\documentclass[runningheads]{llncs}

\usepackage{graphicx}
\usepackage[colorlinks]{hyperref}
\usepackage{caption}
\usepackage{subcaption}
\usepackage[firstpage]{draftwatermark}

\PassOptionsToPackage{usenames,dvipsnames}{xcolor}
\usepackage{xcolor}
\usepackage[utf8]{inputenc} %useful to type directly diacritic characters
\usepackage[procnames]{listings}
\usepackage{lstautogobble}
\usepackage{algpseudocode}
\definecolor{keyword}{HTML}{6852BB}
\definecolor{number}{HTML}{4042B3}
\definecolor{instance}{HTML}{3C8031}
\definecolor{specifier}{HTML}{146666}
\definecolor{operator}{HTML}{C7161C}
\definecolor{operatorword}{HTML}{992024}
\definecolor{functioncall}{HTML}{1773cf}
\definecolor{magicname}{HTML}{B32483}
\definecolor{property}{HTML}{B34024}
\begin{document}
\title{ 3D Environment Modeling for Falsification and Beyond with Scenic 3.0}
\author{Eric Vin\inst{1}\orcidID{0000-0002-3089-1129} \and Shun Kashiwa\inst{1} \and Matthew Rhea\inst{3} \and Daniel J. Fremont\inst{1}\orcidID{0000-0002-9992-9965} \and \\Edward Kim\inst{2} \and Tommaso Dreossi\inst{4} \and Shromona Ghosh\inst{5} \and Xiangyu Yue\inst{6} \and \\Alberto L. Sangiovanni-Vincentelli\inst{2} \and Sanjit A. Seshia\inst{2}}
\authorrunning{Vin et al.}
\institute{University of California, Santa Cruz
\and University of California, Berkeley
\and SentinelOne
\and insitro
\and Waymo LLC
\and The Chinese University of Hong Kong\\
\email{\{evin, shkashiw, dfremont\}@ucsc.edu}
}
\maketitle              % typeset the header of the contribution
\begin{abstract}
We present a major new version of Scenic, a probabilistic programming language for writing formal models of the environments of cyber-physical systems.
Scenic has been successfully used for the design and analysis of CPS in a variety of domains, but earlier versions are limited to environments that are essentially two-dimensional.
In this paper, we extend Scenic with native support for 3D geometry, introducing new syntax that provides expressive ways to describe 3D configurations while preserving the simplicity and readability of the language.
We replace Scenic's simplistic representation of objects as boxes with precise modeling of complex shapes, including a ray tracing-based visibility system that accounts for object occlusion.
We also extend the language to support arbitrary temporal requirements expressed in LTL, and build an extensible Scenic parser generated from a formal grammar of the language.
Finally, we illustrate the new application domains these features enable with case studies that would have been impossible to accurately model in Scenic 2.

\end{abstract}

\SetWatermarkAngle{0}
\SetWatermarkText{\raisebox{10cm}{%
  \hspace{0.1cm}%
  \href{https://doi.org/10.5281/zenodo.7887049}{\includegraphics{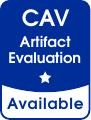}}% 
  \hspace{9cm}%
  \includegraphics{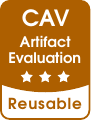}%
}}

\section{Introduction}

A major challenge in the design of cyber-physical systems (CPS) like autonomous vehicles is the heterogeneity and complexity of their environments.
Increasingly, problems of perception, planning, and control in such environments have been tackled using machine learning (ML) algorithms whose behavior is not well-understood.
This trend calls for verification techniques for ML-based CPS; however, a significant barrier has been the difficulty of constructing \emph{formal models} that capture the diversity of these systems' environments~\cite{seshia2016}.
Indeed, building such models is a prerequisite not only for verification but any formal analysis.

Scenic~\cite{scenic_fremont_2019,scenic_fremont_2022} is a probabilistic programming language that addresses this challenge by providing a precise yet readable formalism for modeling the environments of CPS.
A Scenic program defines a \textit{scenario} describing physical objects in a world, placing a probability distribution on their positions and other properties; a single program can generate many different concrete \textit{scenes} by sampling from this distribution.
Scenic also allows defining a stochastic policy describing how agents behave over time, and implementing the resulting dynamic scenarios in a variety of external simulators.
Environment models defined in Scenic can be used for many tasks: falsification, as in the VerifAI toolkit~\cite{verifai_dreossi_2019}, but also debugging, training data generation, and real-world experiment design~\cite{fremont-itsc20}.
These tasks have been successfully demonstrated in a variety of domains including autonomous driving~\cite{av_viswanadha_2021}, aviation~\cite{boeing_fremont_2020}, and reinforcement learning agents~\cite{rl_azad_2022}.

Despite Scenic's successes, it has several limitations that prevent its use in a number of applications of interest.
First, the original language models the world as being \emph{two-dimensional}, since this enables a substantial simplification in the language's syntax (e.g., orientations being a single angle) as well as optimizations in its implementation.
The 2D assumption is reasonable for domains such as driving but leaves Scenic unable to properly model environments for aerial and underwater vehicles, for example.
There can be problems even for ground vehicles: Scenic could not generate a scene where a robot vacuum is underneath a table, as their 2D bounding boxes would overlap and Scenic would treat them as colliding.
The use of bounding boxes rather than precise shapes also leads Scenic to use a simplistic visibility model that ignores occlusion, making it possible for Scenic to claim objects are visible when they are not and vice versa: a serious problem when generating training data for a perception system.

Fundamentally, verification of AI-based autonomous systems requires reasoning about perception and physics in a 3D world.
To support such reasoning, a formal environment modeling language must provide faithful representations of 3D geometry.
Towards this end, we present Scenic 3.0 \footnote{Available at: \url{https://github.com/BerkeleyLearnVerify/Scenic/}}, a largely backwards-compatible major release featuring:
\begin{itemize}
    \item \textbf{Native 3D Syntax}: We update Scenic's existing syntax to support 3D geometry, and add new syntax making it possible to define complex 3D scenarios simply. For example, an object's orientation can be specified as being tangent to a surface and facing another object as much as possible.
    \item \textbf{Precise 3D Shapes}: The shapes of objects (as well as surfaces and volumes) can be given by arbitrary 3D meshes, with Scenic performing precise reasoning about collisions, containment, tangency, etc.
    \item \textbf{Precise Visibility}: We use ray tracing for precise visibility checks that take occlusion into account.
    \item \textbf{Temporal Requirements}: We support arbitrary Linear Temporal Logic~\cite{ltl_amir_1977} properties to constrain dynamic scenarios (vs. only $\textbf{G} p$ and $\textbf{F} p$ in Scenic 2).
    \item \textbf{Rewritten Parser}: We give a Parsing Expression Grammar~\cite{peg_ford_2004} for Scenic, using it to generate a parser with more precise error messages and better support for new syntax and optimization passes.
\end{itemize}

We first define the new features in Scenic 3 in detail in Sec.~\ref{section:new_features}, working through several toy examples.
Then, in Sec.~\ref{section:examples}, we describe two case studies using Scenic with scenarios that could not be accurately modeled without the new features: falsifying a specification for a robot vacuum and generating training data constrained by an LTL formula for a self-driving car's perception system.

\paragraph{Related Work.}
There are many tools for test and data generation~\cite{testing_broy_2005}.
Some approaches learn from examples~\cite{fuzzing_sutton_2007,synthesis_fisher_2012} and so do not provide specific control over scenarios as Scenic does.
Approaches based on rules or grammars~\cite{fuzzing_sutton_2007,modeling_muller_2006,synthesis_jiang_2018} provide some control but have difficulty enforcing requirements over the generated data as a whole.
Several probabilistic programming languages have been used for generation of objects and scenes~\cite{dippl_stuhlmuller_2014,quicksand_ritchie_2014,thesis_ritchie_2014}, but none of them provide specialized syntax to lay out geometric scenarios, nor for describing dynamic behaviors.
Finally, there has been work on synthetic data generation of 3D scenes and objects using ML techniques such as GANs (e.g., \cite{synthesis_fisher_2012,synthesis_wang_2018,shapecrafter_fu_2022}), but these lack the specificity and controllability provided by a programming language like Scenic.

\section{New Features}
\label{section:new_features}

\subsection{3D Geometry}
\label{feature:3d_geometry}

The primary new feature in Scenic 3 is the generalization of the language to 3 dimensions.
Some changes, like changing the type system so that vectors have length 3, are obvious: here we focus on cases where the existing syntax of Scenic does not easily generalize, using simple scenarios to motivate our design choices.

The first challenge when moving to 3D is the representation of an object's orientation in space: Scenic's existing \lstinline{heading} property, providing a single angle, is no longer sufficient.
Instead, we introduce \lstinline{yaw}, \lstinline{pitch}, and \lstinline{roll} angles, using the common convention for aircraft that these represent \emph{intrinsic} rotations (i.e., \lstinline{yaw} is applied first, then \lstinline{pitch} is applied to the resulting orientation, etc.).
Using intrinsic angles makes it easy to compose rotations: for example if we point an airplane towards a landing strip with \lstinline{yaw} and \lstinline{pitch} (either manually or using Scenic's \lstinline{facing toward} specifier --- more on this below), we can add an additional \lstinline{roll} by adding to that property.
To further simplify composition, we add a \lstinline{parentOrientation} property which specifies the local coordinate system in which the 3 angles above should be interpreted (by default, the global coordinate system).
This allows the user to specify an orientation with respect to a previously-computed orientation, for instance that of a tilted surface.

Scenic provides a flexible system of natural language \textit{specifiers} which can be combined to define properties of objects.
Consider the following Scenic 3 code:

\begin{minipage}{\linewidth}
\begin{lstlisting}
objectA = new Object at (1, 2, 3), facing (45 deg, 0, 90 deg)
objectB = new Object left of objectA by 1
objectC = new Object above objectB by 1,
    facing (Range(0,30) deg, Range(0,30) deg, 0)
\end{lstlisting}
\end{minipage}

Here, we use the \lstinline{at} specifier to define a specific \lstinline{position} for object A; the \lstinline{facing} specifier defines the object's \lstinline{orientation} using explicit yaw, pitch, and roll angles.
We then place object B left of A by 1 unit with the \lstinline{left of} specifier: this specifier now not only sets the \lstinline{position} property, but also sets the \lstinline{parentOrientation} property to the orientation of object A (unless explicitly overridden).
Thus object B will be oriented the same way as A.
Similarly, object C is positioned relative to B and so inherits its orientation as its \lstinline{parentOrientation}.
However, this time we use the \lstinline{facing} specifier to define random \lstinline{yaw} and \lstinline{pitch} angles, so object C will face up to $30^\circ$ off of B.

\begin{figure}[tb]
    \centering
    \begin{lstlisting}
    ego = new Ball at (0,0, 1.25)
    new Plane at (2,0,0), facing toward ego
    new Plane at (-2,0,0), facing directly toward ego
    \end{lstlisting}
    \includegraphics[width=0.7\textwidth]{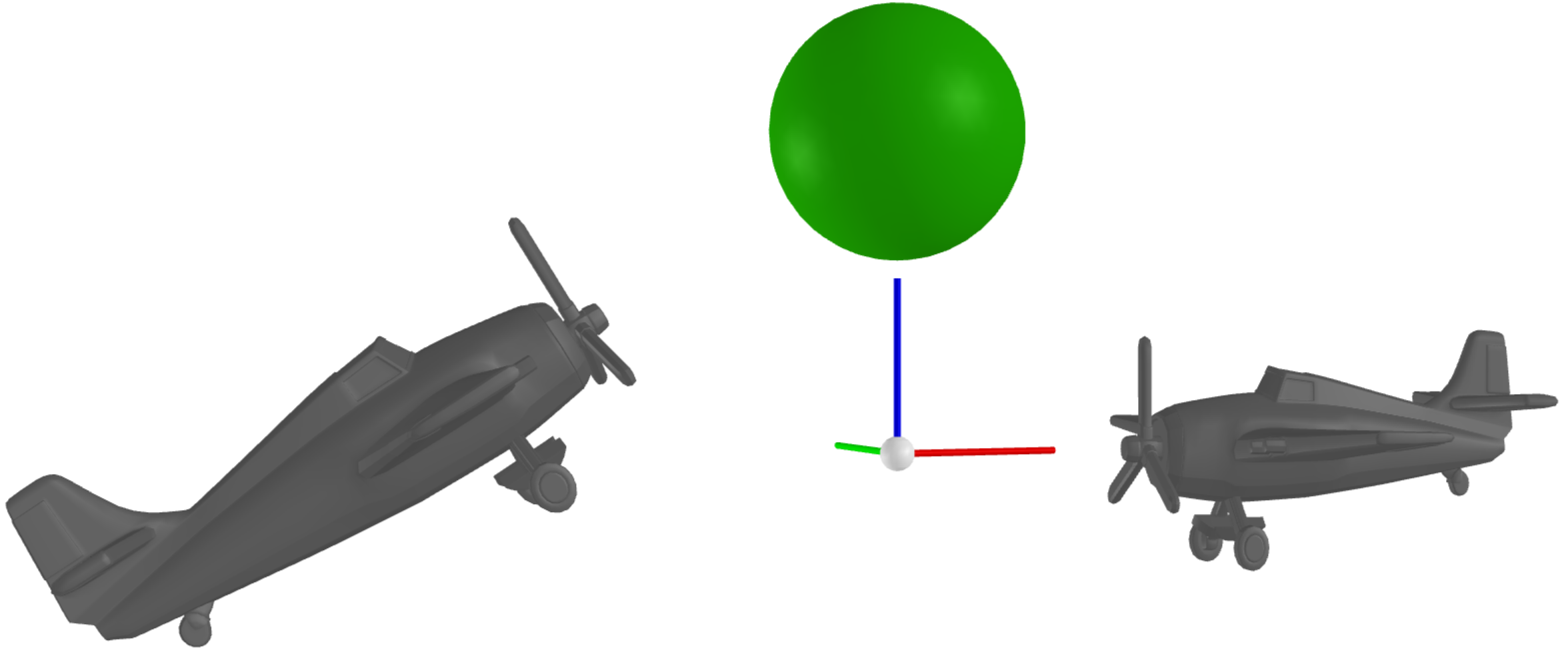}
    \caption{Line-of-sight-based orientations in Scenic. The ego ball (highlighted green) is placed above the origin, as seen by the RGB global coordinate axes, with one plane facing towards the ego and another facing directly toward the ego.}
    \label{scenic:facing}
\end{figure}

Another way to specify an object's orientation is the \lstinline{facing toward} specifier.
This is a case where the 2D semantics become ambiguous in 3D.
Consider a scenario where the user wants an airplane to be ``facing toward'' a runway: the plane's body should be oriented toward the runway (giving its yaw), but it is not clear whether in addition the plane should be pitched downward so that its nose points directly toward the runway.
To allow for both interpretations, Scenic 3 has \lstinline{facing toward} only specify \lstinline{yaw}, while the new \lstinline{facing directly toward} specifier also specifies \lstinline{pitch}.
This is illustrated in Fig.~\ref{scenic:facing}.

\begin{figure}[tb]
    \centering
    \begin{lstlisting}
    floor = Object with width 5, with length 5, with height 0.1
    ego = new Chair on floor
    \end{lstlisting}
    \includegraphics[width=0.7\textwidth]{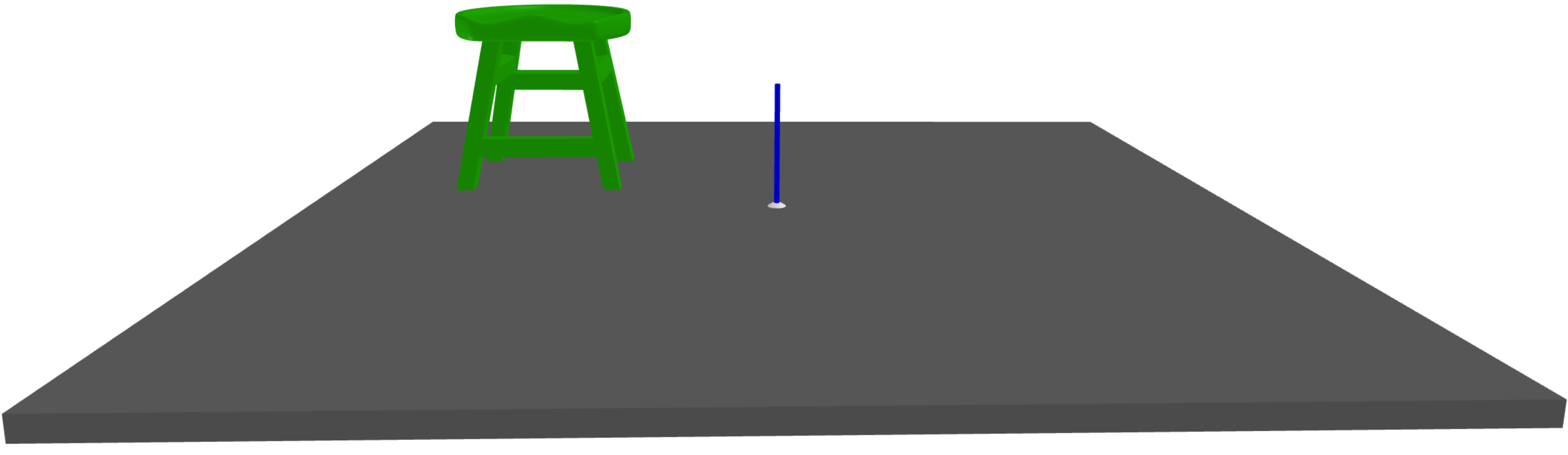}
    \caption{A Scenic program placing a chair on a floor. The Z-axis of the global coordinate axes protrudes from the floor, indicating which direction is up.}
    \label{scenic:on}
\end{figure}

Another common practice in 3D space is to place one object \textit{on} another.
For example, we may want to place a chair on a floor, or a painting on a wall.
Scenic's existing \lstinline{on} specifier, which sets the \lstinline{position} of an object to be a uniformly random point in a given region, does not suffice for such cases because it would cause the chair to intersect the floor or the painting to penetrate the wall (or both).
To fix this issue, we allow each object to define a \emph{base} point, which \lstinline{on} positions instead of the object's center.
The default base point is the bottom center of the object's bounding box, suitable for cars and chairs for example; a \texttt{Painting} class could override this to be the back center.
Finally, to enable placing objects on each other, objects can provide a \lstinline{topSurface} property specifying the surface which is considered the ``top'' for the purposes of the \lstinline{on} specifier.
As before, there is a reasonable default (the upward-pointing faces of the object's mesh) that can be overridden.
This syntax is illustrated in Fig.~\ref{scenic:on}.

\begin{figure}[tb]
    \centering
    \begin{lstlisting}
    floor = new Object with width 5, with length 5, with height 0.1
    air_cube = new Object at (Range(-5,5), Range(-5,5), 3), 
    	facing (Range(0,360 deg), Range(0,30 deg), 0)
    new Chair below air_cube, with color (0,0,200)    # blue chair
    ego = new Chair below air_cube, on floor    # green chair
    \end{lstlisting}
    \includegraphics[width=0.7\textwidth]{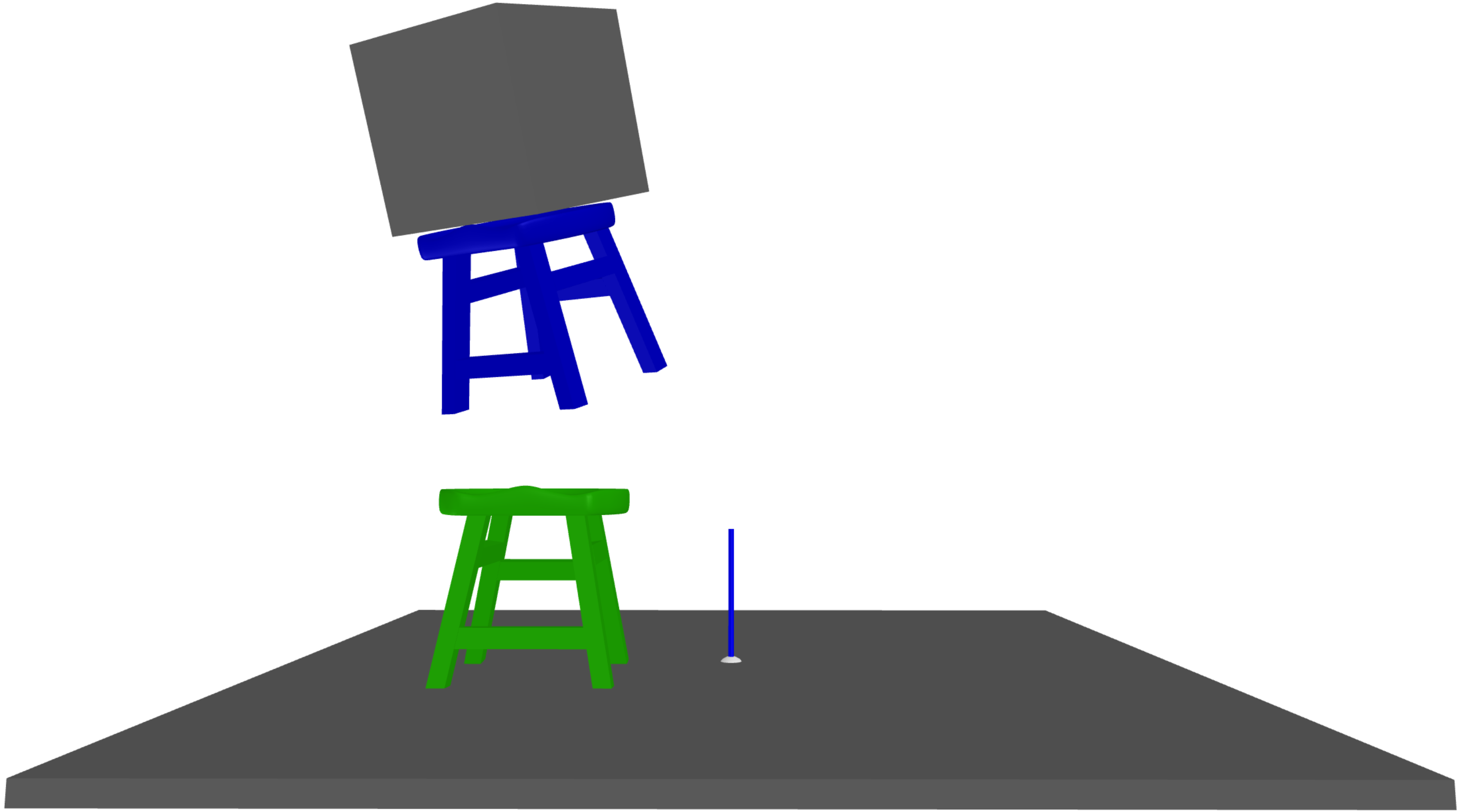}
    \caption{A Scenic program placing a green chair on the floor under a rotated cube in midair. A blue chair is placed directly under the cube for clarity.}
    \label{scenic:on_modifying}
\end{figure}

A final 3D complication arises when positioning objects on irregular surfaces.
Consider a pair of cars driving up an uneven mountain road, with one 10 meters behind the other.
We can use the \lstinline{ahead of} specifier to place one car 10 meters ahead of the other, but then the car will penetrate the road due to its upward slope.
Alternatively, the \lstinline{on} specifier can correctly place the car so it is tangent to the road, but then we cannot directly specify the distance between the cars.
The natural semantics here would be to combine the constraints from \emph{both} specifiers, but this is illegal in Scenic 2 where a given property (such as \lstinline{position}) can only be specified by a single specifier at a time.
We enable this usage in Scenic 3 by introducing the concept of a \emph{modifying specifier} that modifies the value of a property already defined by another specifier.
Specifically, if an object's \lstinline{position} is already specified, the \lstinline{on} specifier will \emph{project} that position down onto the given surface.
This is illustrated by the green chair in Fig.~\ref{scenic:on_modifying}.

Note that the green chair is correctly upright on the floor even though it was positioned relative to the cube, and so should inherit \lstinline{parentOrientation} from the cube as discussed above.
In this situation, the user has provided no explicit orientation for the chair, and both \lstinline{below} and \lstinline{on} can provide one.
To resolve this ambiguity, we introduce a \emph{specifier priority} system, where specifiers have different priorities for the properties they specify (generalizing Scenic's existing system where a specifier could specify a property \emph{optionally}).
In our example, \lstinline{below} specifies \lstinline{position} with priority 1 and \lstinline{parentOrientation} with priority 3, while \lstinline{on} specifies these with priorities 1 and 2 respectively.
So both specifiers determine \lstinline{position} (with \lstinline{on} modifying the value from \lstinline{below} as explained above), but \lstinline{on} takes precedence over \lstinline{below} when specifying \lstinline{parentOrientation}.
This yields the expected behavior while still allowing \lstinline{below} to determine the orientation when used in combination with other specifiers than \lstinline{on}.

\subsection{Mesh Shapes and Regions}
\label{feature:mesh_shapes_regions}

Scenic 2's approximation of objects by their bounding boxes was adequate for 2D driving scenarios, for example, but is wholly inadequate in 3D, where objects are commonly far from box-shaped.
For example, consider placing a chair tucked in under a table.
Since the bounding boxes of these two objects intersect, Scenic 2 would always reject this situation as a collision and try to generate a new scene, even if the chair and table are entirely separate.
In Scenic 3, each object has a precise shape given by its \lstinline{shape} property, which is set to an instance of the class \texttt{Shape}.
The most general \texttt{Shape} class is \texttt{MeshShape}, which represents an arbitrary 3D mesh and can be loaded from standard formats; classes for primitive shapes like spheres are provided for convenience.
These shapes are used to perform precise collision and containment checks between objects and regions.

Scenic also supports mesh regions, which can either represent surfaces or volumes in 3D space. For example, given a mesh representing an ocean we might want to sample on the surface for a boat or in the volume for a submarine.

All meshes in Scenic are handled using Trimesh \cite{trimesh}, a Python library for triangular meshes, which internally calls out to the tools Blender \cite{blender} and OpenSCAD \cite{openscad} for several operations.
These operations tend to be expensive, so Scenic uses several heuristics to cheaply determine simple cases; these can give between a 10x-1000x speedup when sampling scenes.

\subsection{Precise Visibility Model}
\label{feature:visibility}

Scenic 2's visibility system simply checks if the bounding box corners of objects are contained in the view cone of the viewing object, which is no longer adequate for 3D scenarios with complex shapes.
Visibility checks are now done using ray tracing, and account for objects being able to occlude visibility.
In addition to standard pyramidal view cones used for cameras, Scenic correctly handles wrap-around view regions such as those of common LiDAR sensors.
Visibility checks use a configurable density of rays, and are optimized to only send rays in areas where they could feasibly hit the object.

\subsection{Temporal Requirements}

A key feature of Scenic is the ability to declaratively impose constraints on generated scenes using \lstinline{require} statements.
However, Scenic 2 only provides limited support for \textit{temporal} requirements constraining how a dynamic scenario evolves over time, with the \lstinline{require always} and \lstinline{require eventually} statements.
Slightly more complex examples, like ``cars A and B enter the intersection after car C'', require the user to explicitly encode them as monitors, which is error-prone and yields verbose hard-to-read imperative code: this property requires an 8-line monitor in \cite{scenic_fremont_2022}.

Scenic 3 extends \lstinline{require} to arbitrary properties in Linear Temporal Logic~\cite{ltl_amir_1977}, allowing natural properties like this to be concisely expressed:
\begin{lstlisting}
require (carA not in intersection and carB not in intersection
    until carC in intersection)
\end{lstlisting}

The semantics of the operators \lstinline{always}, \lstinline{eventually}, \lstinline{next}, and \lstinline{until} are taken from RV-LTL~\cite{ltl_semantics_bauer_2010} to properly model the finite length of Scenic simulations.

\subsection{Rewritten Parser}
\label{feature:parser}

For interoperability with Python libraries, Scenic is compiled to Python, and the original Scenic parser was implemented on top of the Python parser.
This approach imposed serious restrictions on the language design (e.g., forcing non-intuitive operator precedences), made extending the parser difficult, and led to misleading error messages which pointed to the wrong part of the program.

Scenic 3 uses a parser automatically generated from a Parsing Expression Grammar (PEG)~\cite{peg_ford_2004} for the language.
The parser is based on Pegen~\cite{pegen}, the parser generator developed for CPython, and the grammar itself was obtained by extending the Python PEG.
The new parser outputs an abstract syntax tree representing the structure of the original Scenic code (unlike the old parser), ensuring that syntax errors are correctly localized and simplifying the task of writing analysis and optimization passes for Scenic.

This new parser gives us flexibility in designing and implementing the language. For example, we carefully assigned precedence to the four new temporal operators so that users can naturally express temporal requirements without unnecessary parentheses.
There are additional benefits from having a precise machine-readable grammar for Scenic: for instance, as we wrote the grammar, we discovered ambiguities that had previously been unnoticed and made minor changes to the language to eliminate them.
The grammar could also be be used to fuzz test the compiler and other tools operating on Scenic programs.

\section{Case Studies}
\label{section:examples}

In this section, we discuss two case studies in the robotics simulator Webots~\cite{webots}. The code for both case studies is available in the Scenic GitHub repository~\cite{scenic_repo}.
The first case study, performing falsification of a robot vacuum, illustrates a domain that could not be modeled in Scenic 2 due to the lack of 3D support.
The second case study, generating data constrained by an LTL formula for testing or training the perception system of an autonomous vehicle, is an example of how the new features in Scenic 3 can significantly improve effectiveness even in one of Scenic's original target domains.

\subsection{Falsification of a Robot Vacuum}
\label{section:examples:vacuum}

\begin{figure}[tb]
    \centering
    \begin{subfigure}[b]{0.49\textwidth}
         \centering
         \includegraphics[width=\textwidth]{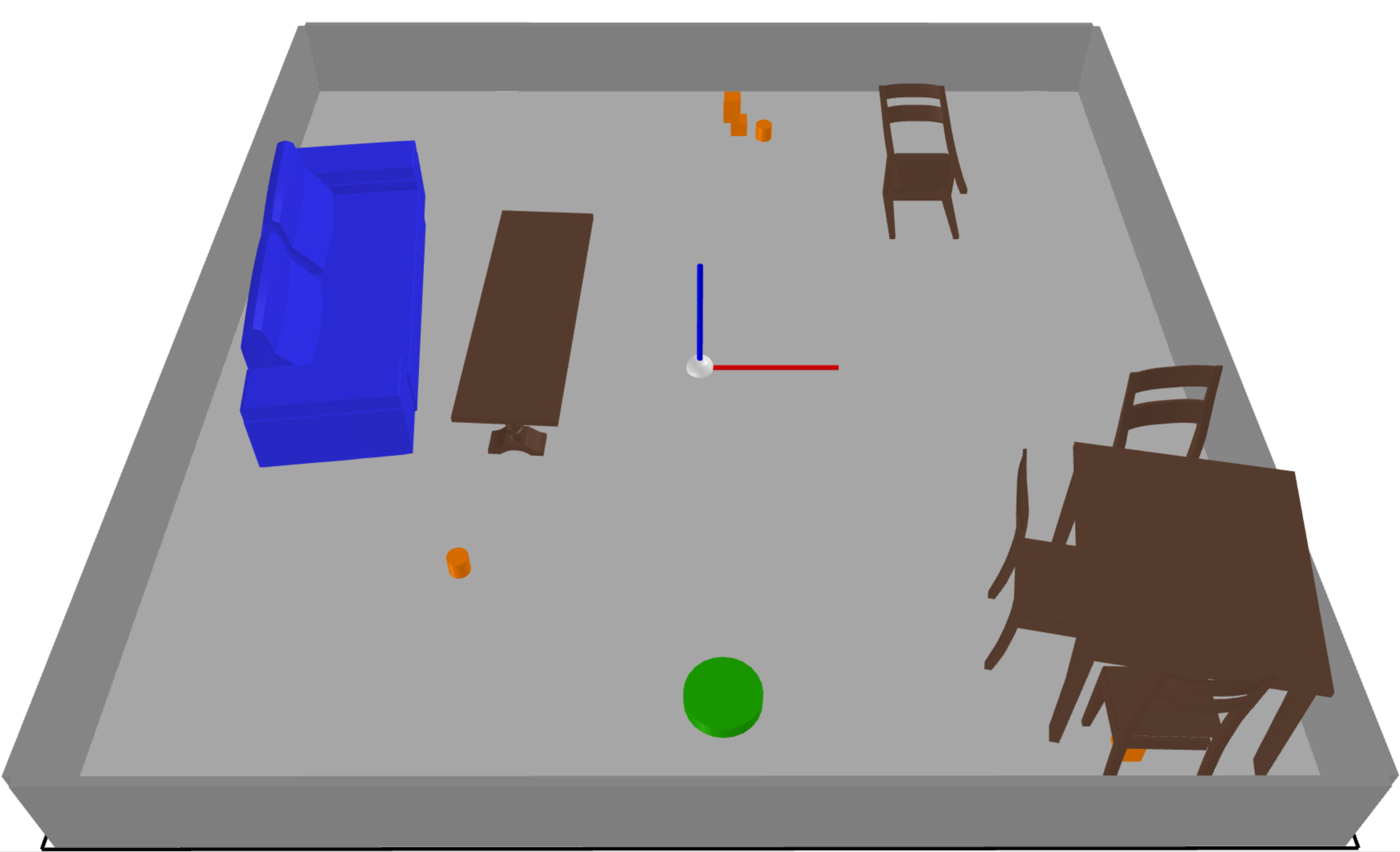}
    \end{subfigure}
    \hfill
    \begin{subfigure}[b]{0.49\textwidth}
         \centering
         \includegraphics[width=\textwidth]{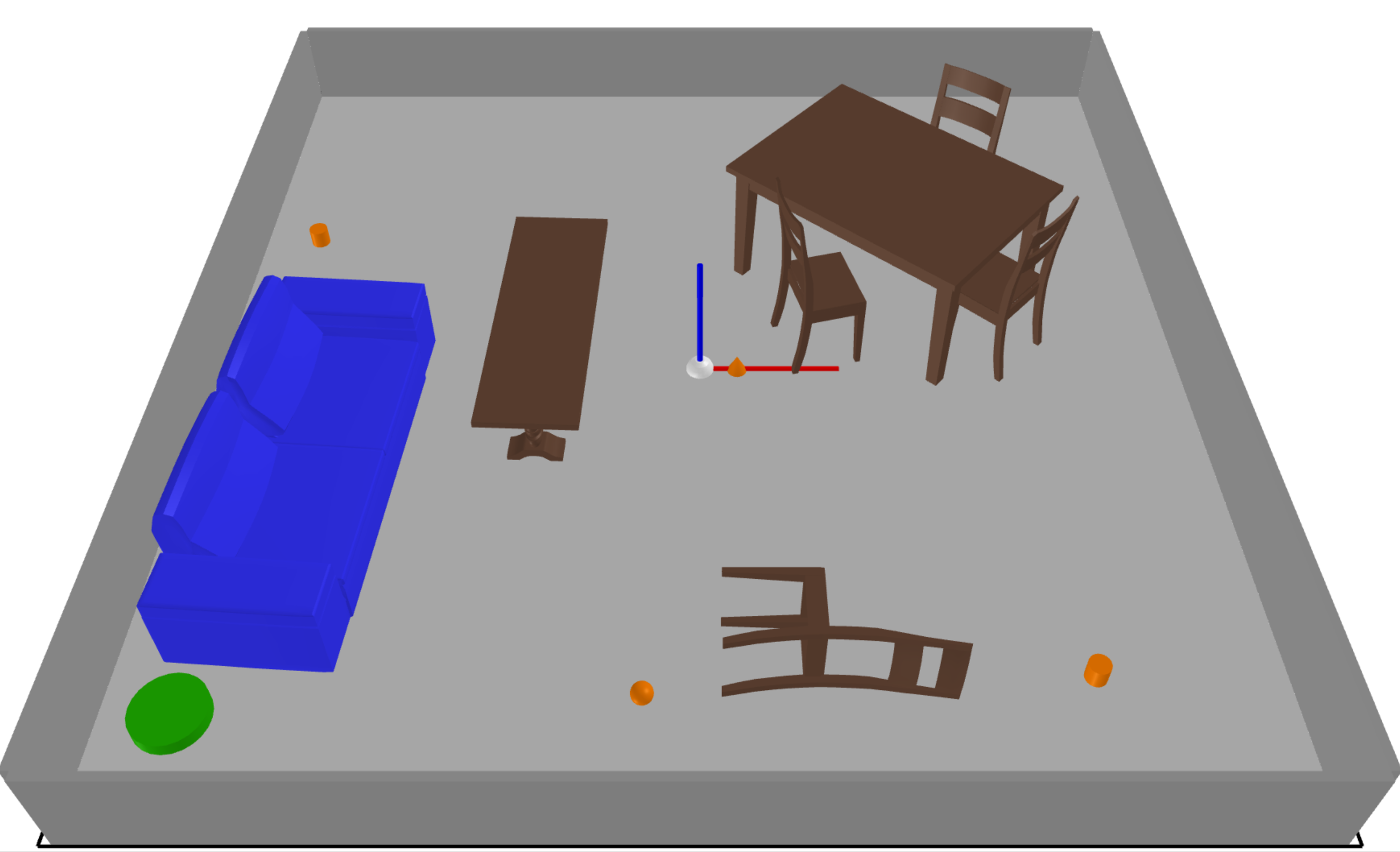}
    \end{subfigure}
    \\
    \begin{subfigure}[b]{0.49\textwidth}
         \centering
         \includegraphics[width=\textwidth]{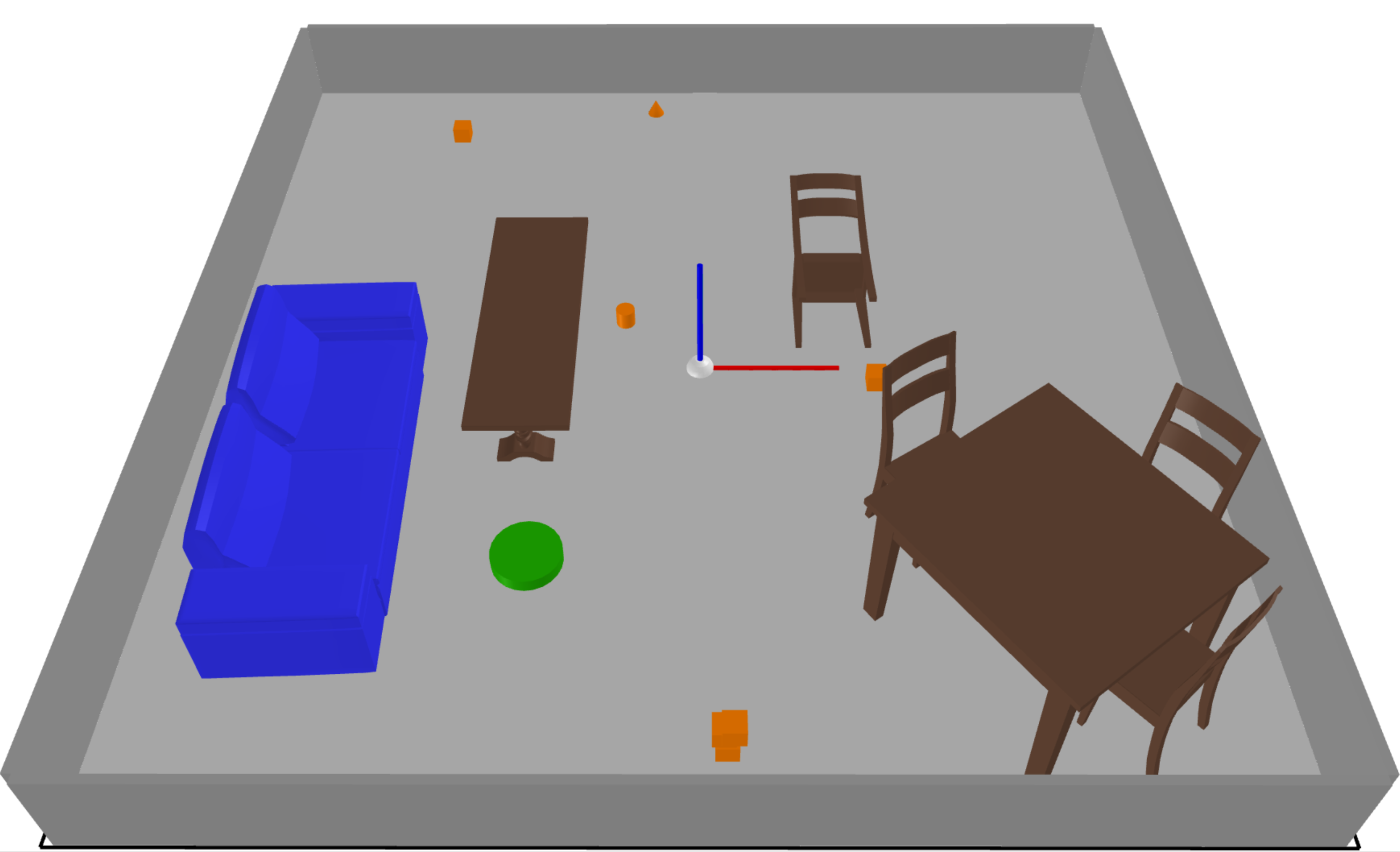}
    \end{subfigure}
    \hfill
    \begin{subfigure}[b]{0.49\textwidth}
         \centering
         \includegraphics[width=\textwidth]{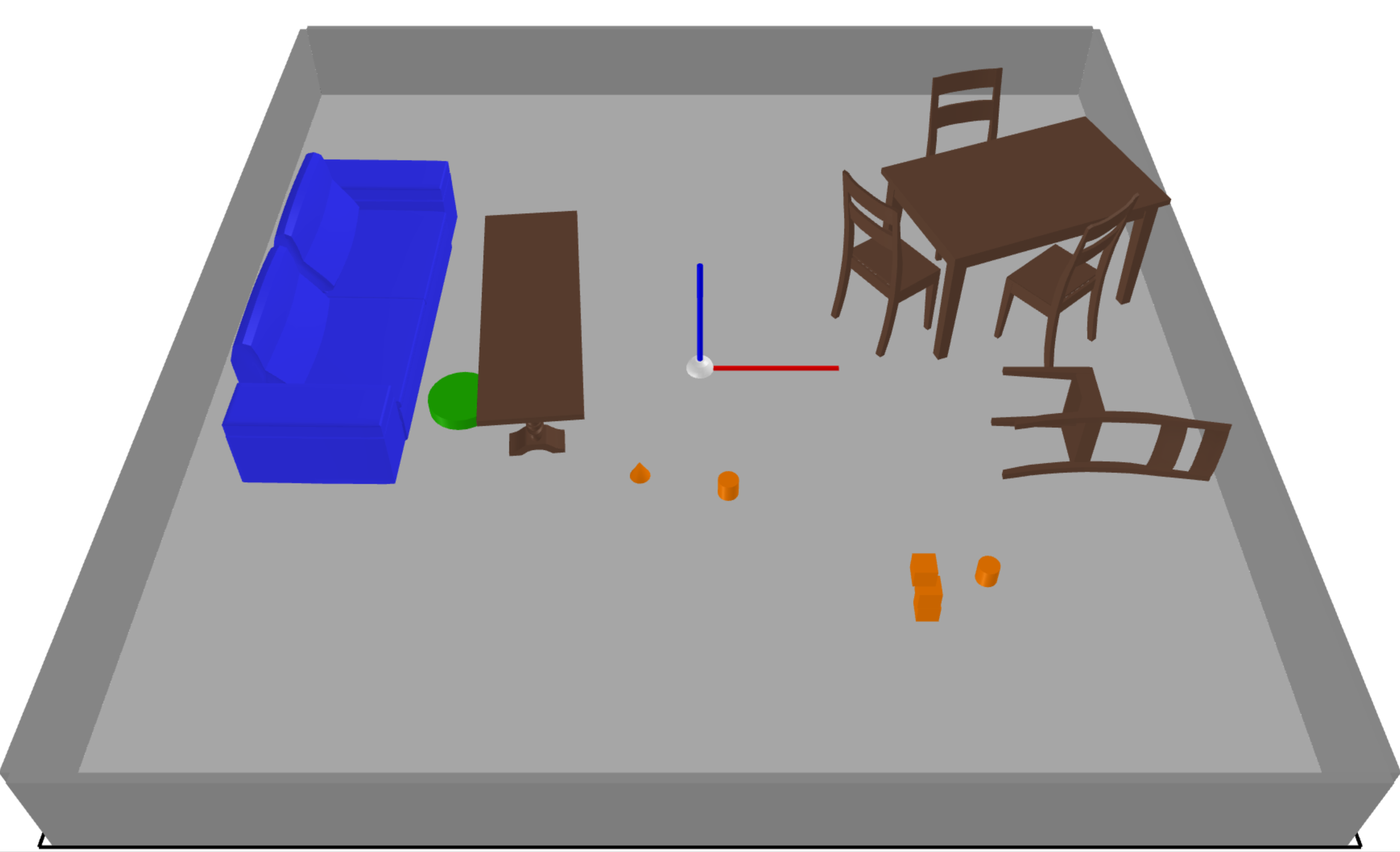}
    \end{subfigure}
    \caption{Several sampled scenes from the robot vacuum scenario.}
    \label{robot_vacuum:example_room}
    \vspace{1em}
    \includegraphics[width=0.8\textwidth]{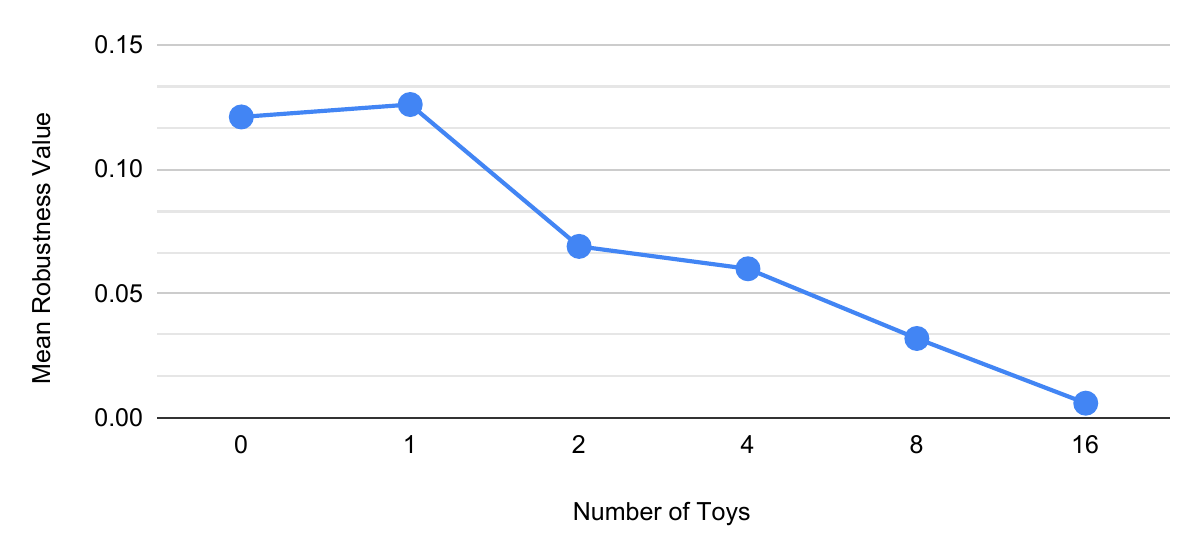}
    \caption{Spec. robustness value vs. number of toys, averaged over 25 simulations.}
    \label{robot_vacuum:coverage_chart}
\end{figure}

In this example we evaluate the iRobot Create~\cite{irobotcreate}, a robot vacuum, on its ability to effectively clean a room filled with objects.
We use a specification stating that the robot must clean at least a third of the room within 5 minutes: in Signal Temporal Logic~\cite{stl}, the formula $\varphi = F_{[0,300]} (\textit{coverage} > 1/3)$.
We use Scenic to generate a complete room and export it to Webots for simulation.
The room is surrounded by four walls and contains two main sections: in the dining room section, we place a table of varied width and length randomly on the floor, with 3 chairs tucked in around it and another chair fallen over.
In the living room section, we place a couch with a coffee table in front of it, both leaving randomly-sized spaces roughly the diameter of the robot vacuum.
We then add a variable number of toys, modeled as small boxes, cylinders, cones, and spheres, placed randomly around the room; for a taller obstacle, we place a stack of 3 box toys somewhere in the room.
Finally, we place the vacuum randomly on the floor, and use Scenic's \lstinline{mutate} statement to add noise to the positions and yaw of the furniture.
Several scenes sampled from this scenario are shown in Fig.~\ref{robot_vacuum:example_room}.

We tested the default controller for the vacuum against 0, 1, 2, 4, 8, and 16-toy variants of our Scenic scenario, running 25 simulations for each variant.
For each simulation, we computed the robustness value~\cite{mtl-robustness} of our spec $\varphi$.
The average values are plotted in Fig.~\ref{robot_vacuum:coverage_chart}, showing a clear decline as the number of toys increases.
Many of the runs actually falsified $\varphi$: up to 44\% with 16 toys.

There are several aspects of this example that would not be possible in Scenic 2.
First, the new syntax in Scenic 3 allows for convenient placement of objects, specifically the use of \lstinline{on} in combination with \lstinline{left of} and \lstinline{right of}, to place the chairs on the appropriate side of the dining table but on the floor.
Many of the objects are also above others and have overlapping bounding boxes, but because Scenic now models shapes precisely, it is able to properly register these objects as non-intersecting and place them in truly feasible locations (e.g., in Fig.~\ref{robot_vacuum:example_room}, the toy under the dining table in the top left scene and the robot under the coffee table in the bottom right scene).

\subsection{Constrained Data Generation for an Autonomous Vehicle}
\label{section:examples:intersection}

In this example we generate instances of a potentially-unsafe driving scenario for use in training or testing the perception system of an AV.
Consider a car passing in front of the AV in an intersection where the AV must yield, and so needs to detect the other car before it becomes too late to brake and avoid a collision.
We want to generate time series of images labeled with whether or not the crossing car is visible, for a variety of different scenes with different city layouts to provide various openings and backdrops.
Our scenario places both the ego car (the AV) and the crossing car randomly on the appropriate road ahead of the intersection.
We place several buildings along the crossing road that block visibility, allowing some randomness in their position and yaw values.
We also place several buildings completely randomly behind the crossing road to provide a diverse backdrop of buildings in the images.
Finally, we want to constrain data generation to instances of this scenario where the crossing car is not visible until it is close to the AV, as these will be the most challenging for the perception system.
Using the new LTL syntax, we simply write:
\begin{lstlisting}
require (not ego can see car) until distance to car < 75
\end{lstlisting}

\begin{figure}[tb]
    \centering
    \begin{subfigure}[b]{0.49\textwidth}
         \centering
         \includegraphics[width=\textwidth, trim=0 400 0 0, clip]{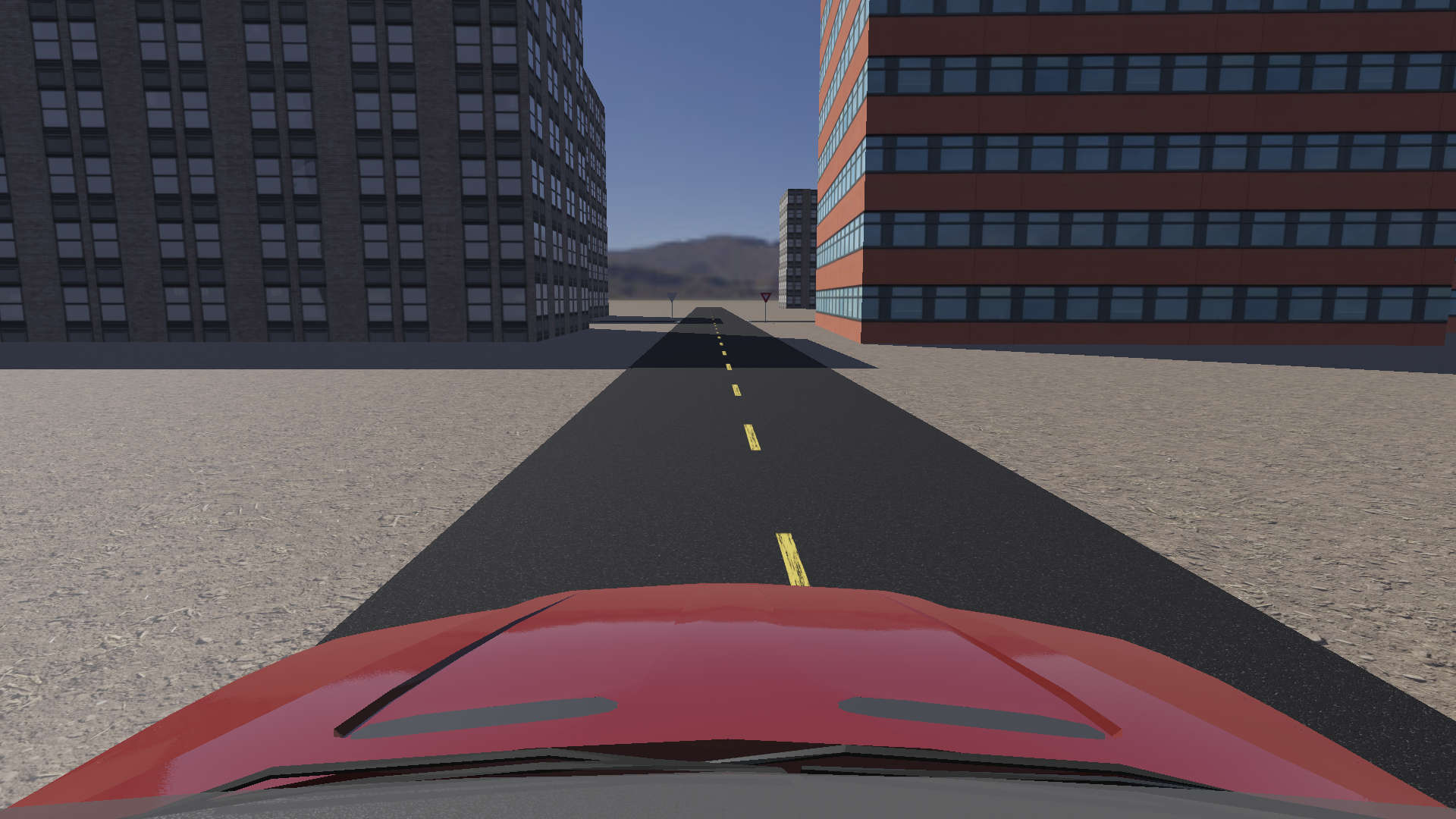}
         \caption{2 seconds: not visible}
    \end{subfigure}
    \hfill
    \begin{subfigure}[b]{0.49\textwidth}
         \centering
         \includegraphics[width=\textwidth, trim=0 400 0 0, clip]{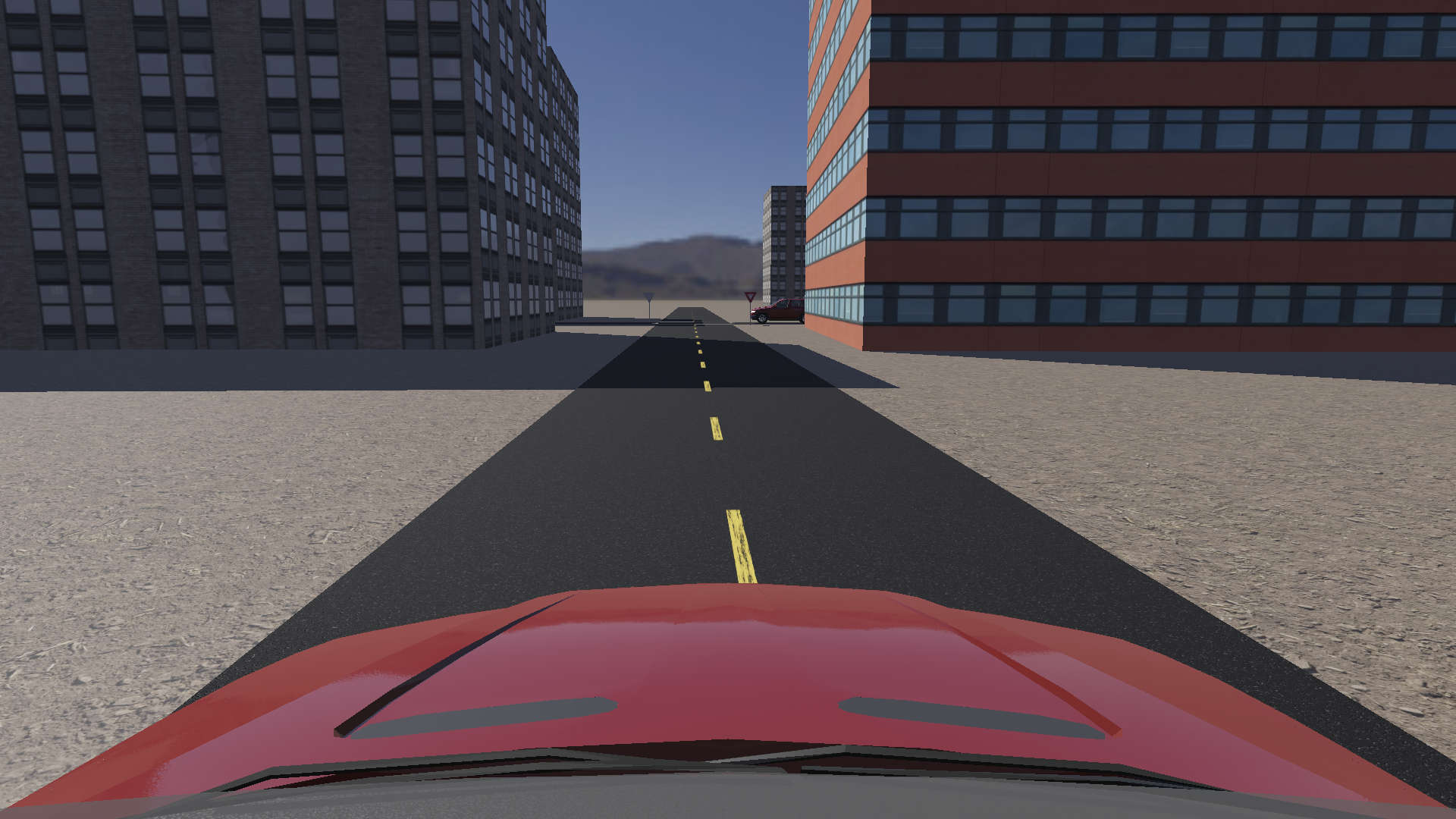}
         \caption{2.5 seconds: visible}
    \end{subfigure}
    \\
    \begin{subfigure}[b]{0.49\textwidth}
         \centering
         \includegraphics[width=\textwidth, trim=0 400 0 0, clip]{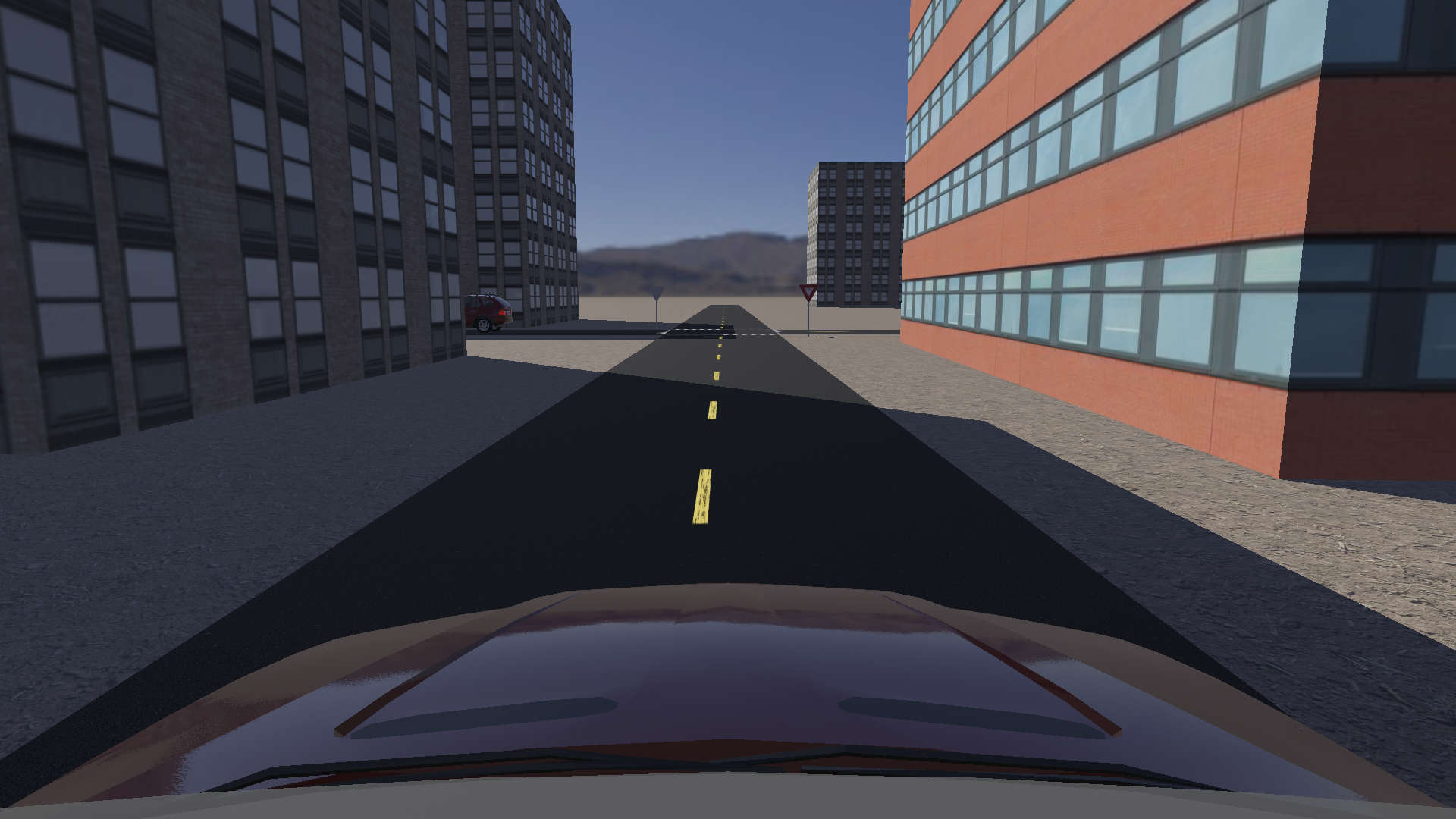}
         \caption{4 seconds: visible}
    \end{subfigure}
    \hfill
    \begin{subfigure}[b]{0.49\textwidth}
         \centering
         \includegraphics[width=\textwidth, trim=0 400 0 0, clip]{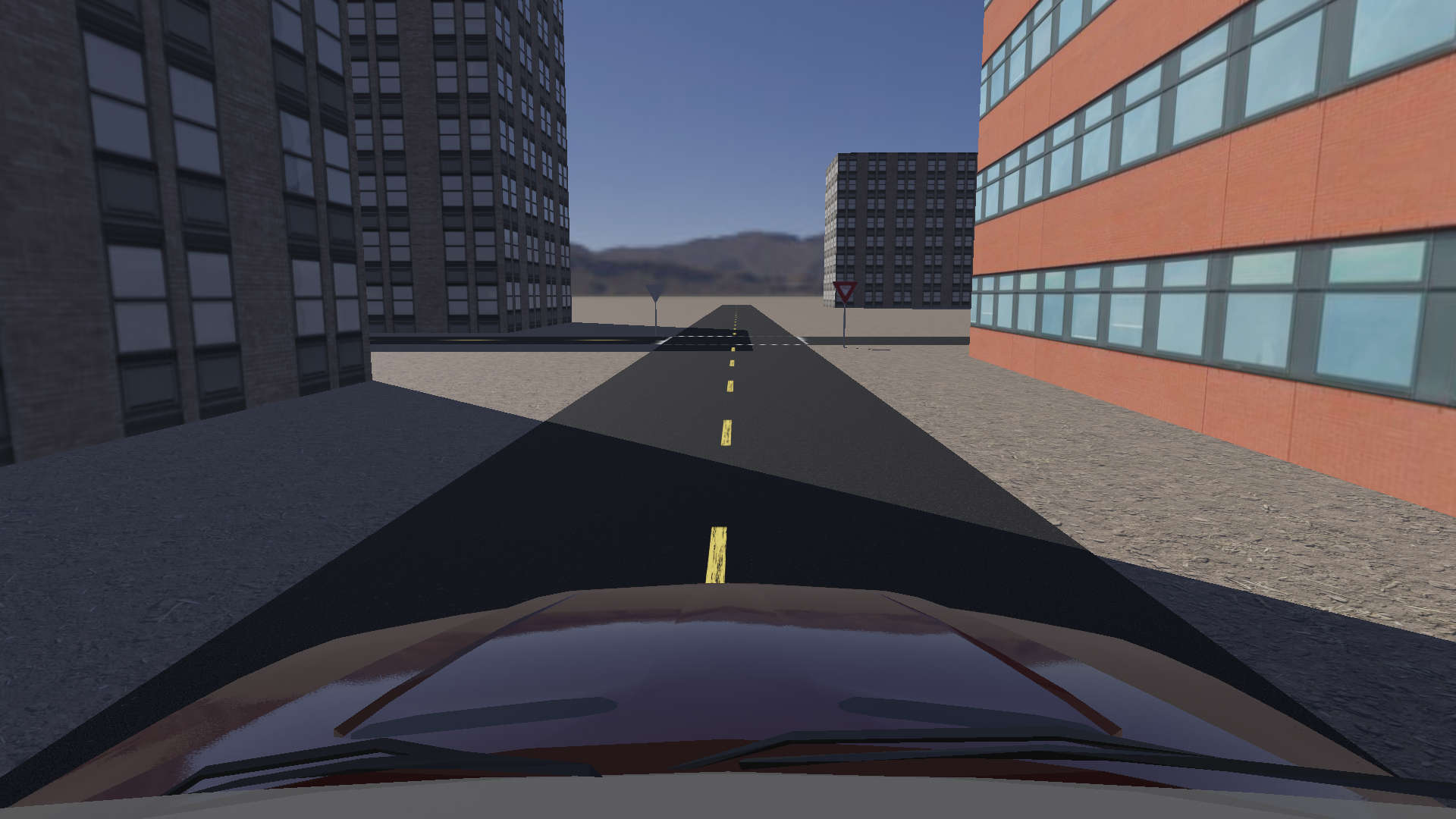}
         \caption{4.5 seconds: not visible}
    \end{subfigure}
    \caption{Intersection simulation images, with visibility label for the crossing car.}
    \label{city_intersection:example_images}
\end{figure}

Fig.~\ref{city_intersection:example_images} shows a simulation sampled from this scenario.
In Scenic 2, the crossing car would be wrongly labeled as visible in image (a), since the occluding buildings would not be taken into account.
This would introduce significant error into the generated training set, which in previous uses of Scenic had to be addressed by manually filtering out spurious images; this is avoided with the new system.

\section{Conclusion}

In this paper we presented Scenic 3, a major new version of the Scenic programming language that provides full native support for 3D geometry, a precise occlusion-aware visibility system, support for more expressive temporal operators, and a rewritten extensible parser.
These new features extend Scenic's use cases for developing, testing, debugging, and verifying cyber-physical systems to a broader range of application domains that could not be accurately modeled in Scenic 2.
Our case study in Section \ref{section:examples:vacuum} demonstrated how Scenic 3 makes it easier to perform falsification for CPS with complex 3D environments.
Our case study in Section \ref{section:examples:intersection} further showed that even in domains that could already be modeled in Scenic 2, like autonomous driving, Scenic 3 allows for significantly more precise specifications due to its ability to reason accurately about 3D orientations, collisions, visibility, etc.; these concepts are often relevant to the properties we seek to prove about a system or an environment we want to specify.
We expect the improvements to Scenic we describe in this paper will impact the formal methods community both by extending Scenic’s proven use cases in simulation-based verification and analysis to a much wider range of application domains, and by providing a 3D environment specification language which is general enough to allow a variety of new CPS verification tools to be built on top of it.

In future work, we plan to develop 3D scenario optimization techniques (complementing the 2D methods Scenic already uses) and explore additional 3D application domains such as drones.
We also plan to leverage the new parser to allow users to define their own custom specifiers and pruning techniques.\\

\noindent\textbf{Acknowledgements}
The authors thank Ellen Kalvan for helping debug and write tests for the prototype, and several anonymous reviewers for their helpful comments.
This work was supported in part by 
DARPA contracts FA8750-16-C0043 (Assured Autonomy) and FA8750-20-C-0156 (Symbiotic Design of Cyber-Physical Systems), by Berkeley Deep Drive, by Toyota through the iCyPhy center, and NSF grants 1545126 (VeHICaL) and 1837132.


\bibliographystyle{splncs04}
\bibliography{refs}

\end{document}